# Model-Based Software Engineering and Ada: Synergy for the Development of Safety-Critical Systems[*]

Andree Blotz, Franz Huber, Heiko Lötzbeyer,
Alexander Pretschner, Oscar Slotosch, Hans-Peter Zängerl

Andree.Blotz@m.eads.net
{loetzbey | pretschn}@in.tum.de
{huber | slotosch | zaengerl}@validas.de

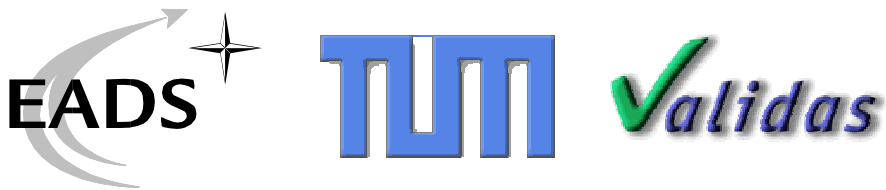

## 1 Introduction

Software development approaches that are based on modeling a system before performing the actual implementation work have a long history in computing. Among the first ones were data(base) modeling approaches using the Entity/Relationship model and similar other techniques. During further development, modeling techniques became increasingly complete, covering not only data aspects, but also structural/topological and behavioral aspects of systems. Typical representatives of such full-scale modeling approaches are structured methods, such as Structured Analysis & Design, or object-oriented methods like the UML [2].

Models created in such a modeling language can serve different purposes. They can be regarded as a concise, much more formal version of otherwise informally given system requirements. In this view, they serve as a precise guideline for the developers that perform the actual implementation work, and can furthermore be used as a basis for testing the conformance of the implementation with the requirements.

If a modeling language is rich enough to allow the creation of *complete* models (models that encompass all aspects of a system on an abstract, implementation-independent level), another purpose of such models is obvious: The created models can not only be used to precisely capture the requirements upon the system, but to describe the system in detail, reaching up to a complete description of all aspects of the system. From such a complete description, it is basically possible (although not always feasible or desired in practice) to generate a complete system implementation automatically. An important advantage of such a model-based approach is (programming) language independence: Modeling languages are usually driven by the application domain that they are used in and provide domain-oriented abstractions to describe systems (components, data entities, states, state transitions, etc.). In contrast, typical programming languages such as Ada or C are general-purpose languages, providing language elements that reflect the underlying machine model of sequential execution of statements. Using code generation techniques to create implementations, such complete models as described previously can be transformed into implementations in arbitrary programming languages.

---

Models are abstractions of a system and are thus particularly less "cluttered" than an implementation, for instance, in C. Therefore, it is much more promising for models than for implementations to apply validation techniques, such as—covering different levels of formality—prototyping and simulation [4], test case/test sequence generation [12], or model checking [7]. If the elements of a modeling language have been chosen carefully enough to keep the modeling language simple, yet complete, it is feasible to provide a sufficiently streamlined formal semantics that even allows the application of rigid formal validation/verification techniques [3].

In contrast, in the development of safety-critical software, processes and quality standards are well-established that are based on the usage of programming languages such as Ada to implement systems, and not on models in arbitrary modeling languages. The MC/DC coverage criterion, for instance, which is required by RTCA/DO-178B [9] for testing safety-critical systems, is defined for programming languages, and not for behavioral models (state diagrams), e.g., in UML or AutoFocus [5].

Thus, the core idea presented in this paper consists in combining the specific strengths and acceptance criteria of both worlds, on the one hand, application domain-oriented, abstract models, and, on the other hand, a well-accepted Ada-based implementation and test setting, by code generation and test case transformation techniques. Another important issue in this context, gaining "trust" into model- and code generation-based techniques can be tackled to a certain extent in this approach as well.

**Structure of this Paper**

This paper is organized as follows: We first introduce a simple, yet powerful modeling language—the AutoFocus modeling language & framework; then we highlight some of the features of our model-based validation toolset. Afterwards, we introduce an example application, the leading edge system (the LES), which is subsequently used to demonstrate a quick walkthrough of our code and test case generation techniques. We sketch how these techniques can be combined in a round-trip fashion into a both model-based and program-based process to benefit from the advantages of both approaches. Finally, we summarize and evaluate our current results and comment on future work to be done in this context.

## 2 AutoFocus Modeling Framework

A modeling language—quite similar to a programming language—comprises a set of concepts that are used to describe systems. In case of programming languages, these concepts are typically statements, blocks, procedures, functions, and many more. For the AutoFocus [5] modeling language and toolset, these concepts are based on the idea of a system being made up of a network of communicating components and are defined in a so-called meta-model (i.e., a model that describes how actual models can be constructed). A simplified representation of the AutoFocus meta-model is shown in Fig. 1, using the UML class diagram notation as the meta-language. The AutoFocus modeling language has been under development at TU München since 1995, specially aimed at the development of embedded systems, and shares some concepts with UML/RT. AutoFocus has been chosen as the modeling platform for the MOBASIS project.

### 2.1 AutoFocus Modeling Concepts

As shown in Fig. 1, the core modeling concepts of AutoFocus, i.e., the core elements in its meta-model are:

**Components.** They are the main building blocks for systems. Components encapsulate *data*, *internal structure*, and *behavior*. Components can communicate with their environment via well-defined interfaces. Components are concurrent: Each one of them runs sequentially; however, in a set of components, each component's run is independent of the other components' runs. A global system clock drives all components in a system, and each component carries out one operation (transition, see below) per system clock cycle. Components can be hierarchically structured, i.e., consist of a set of communicating sub-components.

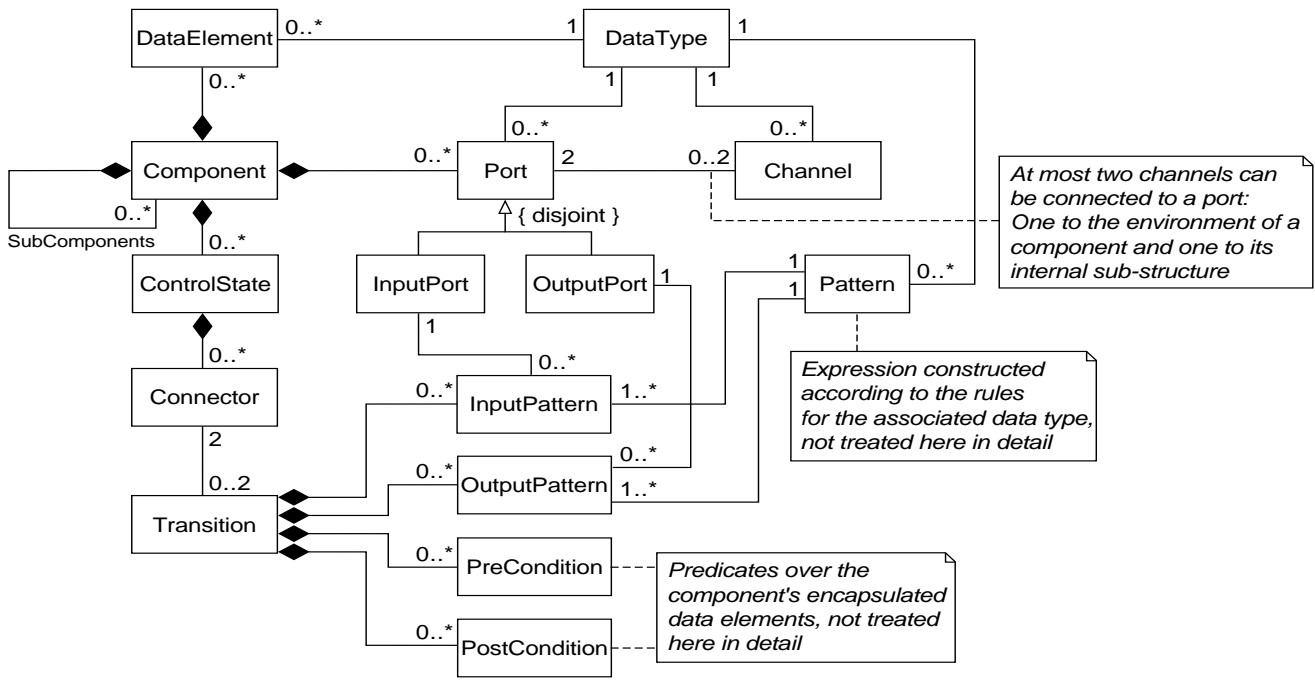

Fig. 1: Basic Modeling Concepts of AutoFocus: The Meta-Model

**Data types.** They define the data structures used by components. Data types are constructed from a set of basic types (such as integer or float) and a set of constructors, e.g., for record and variant types.

**Data.** Data elements are encapsulated by a component and provide a means to store persistent state information inside a component. Data elements can be regarded as typed state variables.

**Ports.** They are a component's means of communicating with its environment. Components read data on input ports and send data on output ports. Ports are named and typed, allowing only specific kinds of values to be sent/received on them. We distinguish two kinds of ports: so-called *immediate* ports, which are visually represented as small, diamond-shaped rectangles, and *delayed* ports (the default variant), which are rendered as small circles. Immediate ports pass along the value that is written to them immediately, that is, within the *same* system clock cycle, whereas delayed ports propagate values written to them not before the *next* system clock cycle.

**Channels.** They connect component ports. Channels are unidirectional, named, and typed, and they define the communication structure (topology) of a system.

**Control States** and **Transitions.** These elements define the control state space and the flow of control inside a component. Each transition connects two distinct controls states (or one control state with itself, in case of a loop transition) and carries a set of four annotations determining its firing conditions (its "enabledness"):

- *guards* and *assignments*, which are predicates over the data elements of the component to be fulfilled before and after the transition, respectively, and
- *input* and *output patterns*, determining which values must be available on the component's input ports to fire the transition and which values are then written to the output ports.

These concepts are sufficient to describe a large class of systems. Developers create the model of an actual system using these concepts; technically speaking, an actual system model is an *instance* of this meta-model. The complete meta-model, together with a set of additional conditions related to consistency and completeness of models, describes the set of all possible, well-formed models that can be created.

## 2.2 AutoFocus Views and Notations

Developers do not create and manipulate models as a whole, but by picking only specific parts of them, which are of interest during particular development activities. These parts, usually closely related with each other, make up the *views* of the system. For instance, the structural view in AutoFocus considers only elements from the meta-model describing the interface of components and their interconnection.

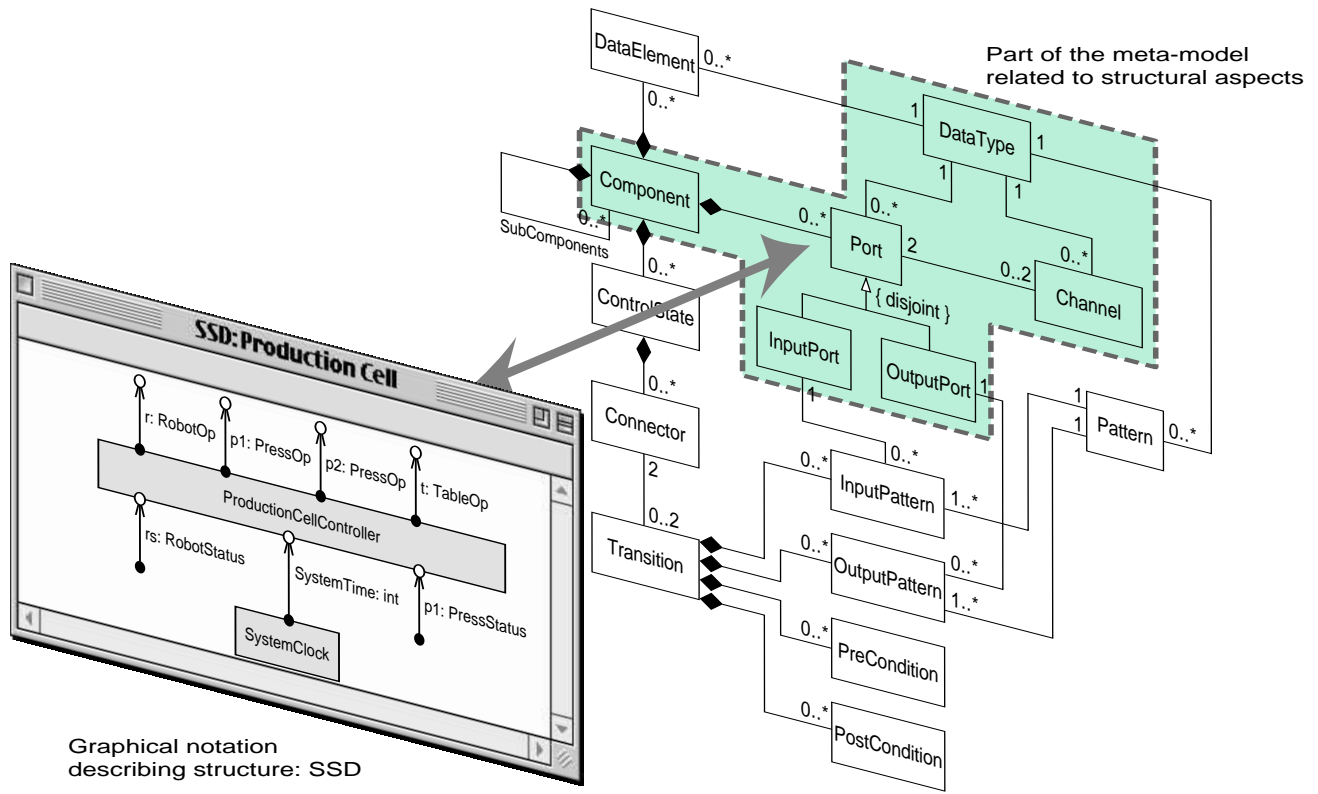

Fig. 2: Structural Parts of the Meta-Model and the Notation representing them

To manipulate elements of these views they must be represented visually to developers. In AutoFocus, mainly graphical notations are used for that purpose; these notations are introduced in more detail by our application example, the LES. The notations do not represent self-contained documents; instead they are a mere visualization of a clipping from the complete model. Fig. 2 shows an example for this relationship between a set of related elements from the meta-model (inside the shaded area) and their visual, diagrammatic representation. In this example, structural aspects of the model are covered, and the notation used to visually represent them is called *System Structure Diagrams* (SSDs). Examples for SSDs are given in the presentation of the LES in Fig. 5 and in Fig. 6. The visual notation used to represent the state-based aspects of a system (control states, transitions) is called *State Transition Diagrams* (STDs) and is shown in Fig. 7. The data type-related aspects of a system are defined using textual *Data Type Definitions* (DTDs); in Section 5.2 we give an example for a DTD. An additional graphical notation used to visualize runs of components over time is called *Extended Event Traces* (EETs) and is similar to UML sequence diagrams [2] and ITU-standardized Message Sequence Charts (MSCs) [6]. EETs play an important role in test case generation. Fig. 8 shows an example of an EET.

## 3  Validation Framework

Validation of models created in AutoFocus is the main goal of the validation framework. A large number of validation approaches and associated tools are available for that purpose. Usually, validation tools (such as model checkers or theorem provers) are standalone tools and are not directly connected to modeling or engineering tools. Therefore, one of the main purposes of the "Validas Validator", the main tool within the validation framework, is to provide a common environment, integrating many commercial and research validation and verification tools by providing generators and translators that allow AutoFocus models to be analyzed by validation tools appropriate for the specific goal of the analysis. Basically, the process works by translating AutoFocus models into the input format of the selected validation tool. Validation results are then retranslated into AutoFocus models after analysis, if this is possible (and desired). In addition to these third-party validation tools, efficient code generators, model metrics and simple consistency checks are part of the Validator. The construction of the validation framework has been initiated by the BSI (Bundesamt für Sicherheit in der Informationstechnik) within the Project Quest [10]. The architecture is open, such that new tools can be easily connected.

### 3.1  Validation Techniques

The available validation techniques cover a broad spectrum, ranging from formal theorem proving to simple and efficient rule checking for consistency checks:

- Theorem proving (VSE Tool)
- Abstraction techniques (Validator)
- Model checking (SMV, NuSMV, Cadence SMV)
- Bounded model checking with propositional solvers (SATO, Chaff)
- Test sequence generation using constraint solvers and backtracking (ECLIPSE)
- Determinism checking (Validator)
- OCL constraint checking (Validator)
- Simulation of executable models (AutoFocus)

For validation of models, sequences (represented as EETs) are the core concept. They can be used as test sequences in several ways: They are generated by some of the aforementioned tools in the framework and can be exported textually for testing the conformance between model and code („regression test'') automatically or to measure the coverage of the code from the generated sequences.

### 3.2  Test Case Generation

Testing and simulation continue to constitute the most popular validation and verification techniques. Their preeminent role in this field is basically due to a lack of serious alternatives: Theorem provers still require intense manual interaction by highly skilled engineers, and the application of push-button approaches like model checking is usually restricted to small or brutally abstracted systems. With suitable ad-hoc abstractions, model checking is successfully deployed for protocol verification, and engineers have been able to find general abstractions in the field of circuit verification—yet, a general abstraction framework for the verification of (hybrid) systems remains to be found. Obviously, the problem is rooted in the fact that abstractions are domain- and application-specific.

In addition to these rather technical issues, verification with theorem provers or model checkers is usually performed on models, or abstractions, of a system, and not on the code. While verifying models is always a useful activity—regardless of whether the models in question have been created before or after the system—validating a system with respect to its requirements has, nonetheless, to be done on the actual system.

For reasons that will be clarified in the sequel, we concentrate on test case specifications (formalizations of a test objective) that are of the form "drive the system into a certain state $q$". Since our focus is on reactive rather than transformative systems, a corresponding test case actually is a test sequence of I/O pairs. The problem of generating test cases, or sequences, respectively, thus amounts to searching a path through a (concurrent) program, and the information one is interested in is finding those I/O pairs that lead to the specified state $q$.

Since we consider the continuous parts of the system to be transformed from differential into difference equations, we get a discrete but infinite state space, and the system may be described by a set of equations. Atomic components translate into an implication of the sort "if the transition's guard holds, then a step from one state to another may be performed". Composed components then consist of a conjunction of all the defining equations for its subcomponents, by turning internal channels into new local variables. This is a (constraint) logic program, and this is almost all we need, both for simulation and test case generation: by giving values to all top level input channels, the whole system performs a step, yielding new values for control and data states, which are then, in turn, used as starting points for a new step of the system, with new inputs. If the guard contains arbitrary equations over reals (which is usually the case for hybrid systems), then we need constraint solvers in addition to logic programming in order to cope with these.

Thus far, we have described a simulation code generator for CLP. In fact, this is already a test case generator, since the concept of input and output variables does not exist in this paradigm—if no values are assigned to some variables, the LP engine computes them. For the generation of test sequences, we partially specify the desired destination state of the system, and make the LP engine do the work. The result of the symbolic execution of the program is, wherever necessary, a binding of all variables, and in particular those variables that occur in the input and output streams. Obviously, the state space usually is far too large for exhaustive exploration, which is why, in addition to elaborate heuristic search algorithms [8], we (automatically) use constraints for a more efficient generation procedure: Parts of the search tree may be pruned in an a-priori manner, shifting the generate-and-test paradigm into constrain-and-generate. If this is still not enough for an efficient computation of test cases, the user can establish environmental and efficiency constraints. Note that we see our approach as complementary and not substituting with respect to model checking: We deliberately drop the completeness requirement.

In any case, we now have described (and actually implemented) a system that is able to find a sequence of I/O pairs that drive a system into our specified state $q$. By comparing actual outputs with desired ones, we can validate models, and with adequate transformations of the test sequences from the level of models to the level of source code, also the system itself, with an automatic assessment of correctness of the system's outputs. This concretization is not a trivial task. So far, we are only able to perform this in an ad-hoc manner, and automatization is the subject of current work. However, drivers for generated Ada code may be generated automatically.

Is finding an I/O sequence leading to $q$ enough for testing systems? Clearly, this kind of "existential" test objectives comprises structural testing, and it is a good debugging aid when designing the model. Furthermore, scenarios from the requirements capture activities also constitute this kind of existential test objectives. However, sometimes "universal" properties like invariance are also established during requirements capture. These properties have the bitter taste of not being testable, since testing activities are finite by definition. For small systems, model checking or theorem proving may be suitable for proving such properties, but again, these techniques can rarely be applied to actual embedded systems. These properties also have to be tested, and they often constitute safety-critical parts of a specification. We will not dive deeply into this, but the idea to cope with this kind of properties is to first negate them. This yields an existential property: There is a path that leads to a state $p$ violating the invariance.

By trying to reach this violating state, we get as "close" as possible to it. Obviously, this requires the notion of a suitable discrete topology, which is difficult to find, but empirically possible [11]. The motivation here is the same as for limit/equivalence class testing: Errors are likely to be found at the boundaries of "intervals", and we define the boundaries to be the "rim" of the set of all reachable states. An empirical assessment of whether or not this is a good idea is the subject of current work.

# 4 Case Study: The Leading Edge System (LES)

The application example chosen for the MOBASIS project is a fictitious Leading Edge System (LES) of a modern fighter jet. It was chosen so as to comprise all the essential features of a modern aerospace flight control system while at the same time it is not diffused by its complexity. Within MOBASIS, this example serves as one of the benchmarks to evaluate the developed techniques and tools for their practical applicability.

## 4.1 Overview of the LES

Leading Edge Systems are so-called secondary control surfaces which are used to increase lift during take-off and landing. They are operated either manually by the pilot or by the autopilot. In instable fighter aircraft the LES is used to stabilize the aircraft along the lateral direction at high incidence angles. Therefore they are permanently used and operated by the Flight Control System (FCS) as a function of incidence angle (AoA) and Mach number. FCS controllers use the pilot reference value from inceptors and measured physical quantities to compute the actuator demands. Now that these demands are safety-critical with failure probabilities less than $10^{-7}..10^{-9}$ per flight hour, the FCS is typically not only designed in a multiple redundant (e.g. quadruplex) way but according to the highest quality assurance and certification measures. The use of Ada with its built-in safety features like, e.g., strong typing and controlled interfaces is therefore mandatory for most aerospace and defense projects.

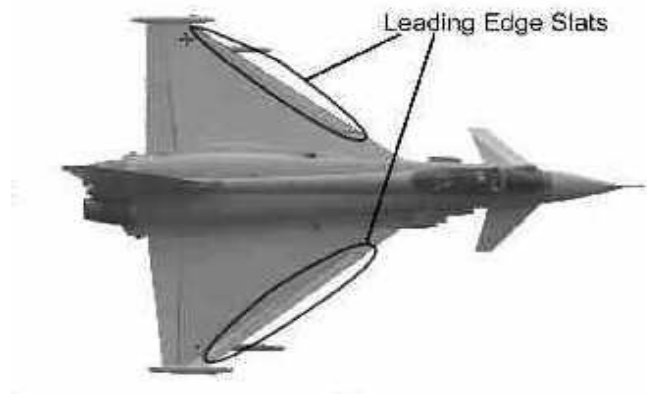

Fig. 3: The Leading Edge System (LES) of an Example Fighter Jet with Slats at the Delta Wing

The LES in this setting therefore implements the following reduced characteristics:

- a sensor voting plane for the incidence angle voting
- a control law calculation including the support of a Ground Support Equipment (GSE)
- a demand voting plane reduced to the LES demand voting
- the actuator loop closure
- the initial built-in test (IBIT)
- a model for the mechanical, hydraulic and electrical part of the LES as far as failure injection is concerned.

The software level criticality of an isolated LES is regarded as mission critical and as software level B according to RTCA/DO-178B. However, since software segregation in the present FCCs is not implemented and the FCCs contain safety critical software of level A type, the whole software within the FCCs has to be developed according to this highest level A.

The redundancy management will be implemented as quadruplex, and a simplified consolidation algorithm calculates the mean value from the two middle values, after appropriate sorting of the four values. In a triplex system or degraded quadruplex system the voter chooses the middle value after sorting.

The sensor voting plane consolidates the input value to the control laws, i.e., the consolidated Angle of Attack (AoA). The AoA is then used by the control law (Fig. 4) to calculate the corresponding LES slats demand. Its function is mainly to provide a smooth position achievement with limited speed, which is implemented via a rate limiter in the control law design. Furthermore, if Ground Support Equipment (GSE) is enabled the actuator demand output is replaced by the analog input from the GSE. On the output side of the Control Law there is another demand voting plane which consolidates the demand before insertion into the slat position loop closure.

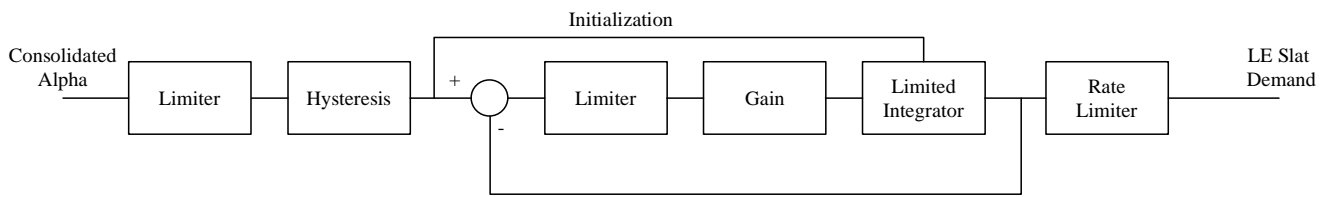

Fig. 4: The Leading Edge Control Law in a Simplified Version

The actuator loop closure then uses the consolidated output to drive electro hydraulic servo valves. These valves drive the main control valves which control the pressurized hydraulic liquid, which itself drives the ram. The rams then are directly coupled to the control surfaces. The control of these hydraulic actuators involves the ram position as well as the position of the valves. The latter control loop is therefore called inner control loop while the former is the outer control loop. Typically, the loop closures are performed at different frequency rates. Both together with the actuators are also called the actuation system. The design of the solenoids must be such that if one of the lanes drives its solenoid against the work of the others that the movement to the true reference position is not prevented. Finally, the build-in test (BIT) performs pre-flight checks, actuator movement checks, and a first line check.

## 4.2 LES-Model in AutoFocus

Summing up the software-related aspects of the previously given description of the LES, three core functional areas can be identified:

- the control laws,
- the quadruplex redundancy control, and
- the self test (IBIT).

Although an extensive self test is required for each safety-critical function, we have decided to concentrate on the other two aspects for modeling the system within the MOBASIS project, as we believe that the self test will influence the complexity of the running system only marginally compared to the influence of a more complex redundancy control. The complete AutoFocus model includes four communicating flight control computers, control laws and a sensor voting plane for voting data from the incidence angle sensor. Due to the complexity of the LES model that has been created in

AutoFocus, and due to the limited space available in this paper, we can only shed light on a few of its most interesting aspects, which are briefly discussed in the following.

The whole LES system consists of three control laws: The LES control law computes the desired slat position, the slat position control law controls actual slat positions and the DDV control law controls the opening and closing of the relevant valves. Fig. 5 illustrates the system structure of the LES control law.

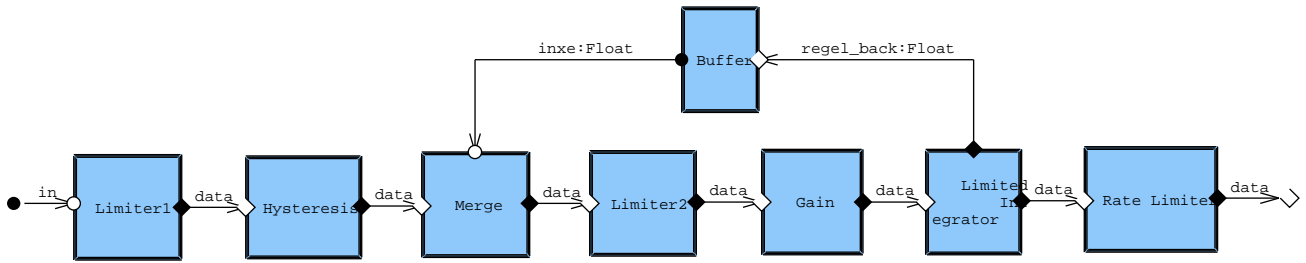

Fig. 5: LES Control Law as an AutoFocus Structure Diagram

The other control laws can be modeled in a similar fashion. Each block ("component") is a function that takes one or two input values and computes an appropriate output value that is transferred immediately (angular ports) to the next block. Therefore, it takes only one cycle for an entering value to be fully processed and emitted to the next control law via the output channel. The only delaying component is the buffer, which stores the output of the limited integrator in the back loop.

In critical FCS applications redundancy control is a major issue. There are different levels of redundancy. For the LES quadruplex redundancy is required. Quadruplex means that the system consists of four FCCs and every sensor signal, intermediate value and actuator demand must be monitored and voted. Fig. 6 shows the monitor/voter plane for the incidence angle sensor signal of one FCC. The "splitter" transfers the value from one of four sensors to the other FCCs and to the monitor. All other sensor values are received indirectly from the other FCCs via the channels "a2" to "a4". The monitor component monitors each signal and decides whether a lane should be activated or deactivated. Internally, the monitor component contains a driver subcomponent for each lane (ports, respectively channels v1 through v4) that activates or deactivates the respective lane.

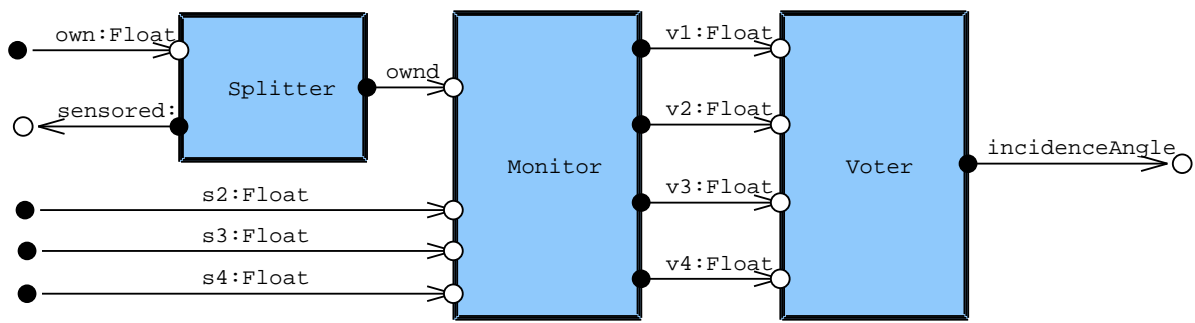

Fig. 6: Monitor/Voter Plane for Incidence Angle Sensor

The state transition diagram shown in Fig. 7 shows the behavioral specification of such a driver component. Note that a lane is neither deactivated nor activated unless a certain number of faulty or correct values have been processed and recorded by the driver, respectively.

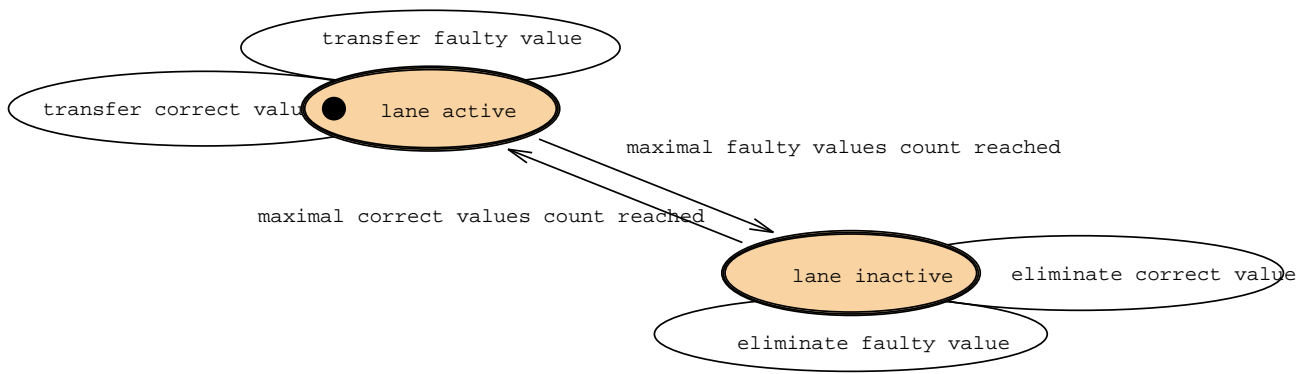

Fig. 7: Specification of a Lane Driver

## 5 From Model to Code

The LES model, of which we have just shown a few aspects, is a *complete* model, which means that it comprises all the necessary information to fully describe each component including its complete behavior. This completeness feature is on the one hand a property of the modeling concepts of AutoFocus: Using all the available concepts in AutoFocus, developers are enabled to create such complete, consistent, and therefore executable models. On the other hand, it is well possible during the development process to create incomplete and contradictory models in AutoFocus. In fact, most of the time during development, models are incomplete, because developers choose to concentrate on specifying only some aspects of a system at a time, such as for instance the structure of a system, leaving open other aspects at the same time, such as data types. AutoFocus provides an extensive set of consistency and completeness checks, which are even user-extensible, to help developers resolve inconsistencies and incompleteness of models. Once a model complies to all the necessary consistency and completeness conditions, it is said to be *executable*, which means that a prototype can be generated and executed within the simulation environment of AutoFocus. Exactly this precondition of executability is also the precondition for the generation of program code from a model. The Validator tool currently provides code generators for C, Java, and—implemented within MOBASIS—for an Ada-subset. The Ada program code currently generated is—in our view—to be used mainly for additional, program code-based validation activities; however, the ultimate goal in code generation is to generate complete system implementations from models. Experiences with our highly optimizing C code generator indicate that this goal should attainable for Ada as well.

### 5.1 Ada for Risk Class 1 Software

The reasons for using Ada, respectively the SPARK, in safety critical projects are twofold. On the non-functional side EADS-M has a long tradition and extensive experience in the development of Ada projects. Also, tools and processes are trimmed to the Ada language and, for some projects, the usage of Ada is part of the customer or government requirements. On the functional side, Ada has many built-in safety features like, e.g., not permitting to read and write outside the bounds of an array, strong typing, modularity, simple syntax etc. Such features enhance reliability by providing readable and maintainable programs. In addition, the SPARK [1] language has removed elements that would prevent a program to be rigorously proven or would tremendously impede the process of verification and validation. This, e.g., would be the case for the effect of goto statements on the control flow analysis. But it should be kept in mind that SPARK is not just a safe subset of Ada but shares a common kernel with Ada and has additional annotations for flow analysis and formal proof. Especially excluded from the kernel are the Ada features for tasking, exceptions, generics and also access types, goto statements and use package clauses. In general, most of the dynamic flexibility that was added to Ada95 had to be excluded since high integrity software does not allow for dynamical processes once the system is initialized, but requires the system to be deterministically defined at start up.

For the MOBASIS Ada code generator, we have chosen to adhere to a project-specific Ada subset from EADS, similar to SPARK, for the first implementation. Future development, however, will use SPARK Ada as reference subset. In the following we briefly describe the generation of Ada code from AutoFocus SSDs, STDs and DTDs (see also Section 2.2).

## 5.2 Ada Code Generation from AutoFocus Models

The data-related aspects of AutoFocus models are based on data type definitions (DTDs) that allow developers to define types, functions and constants in a fashion very similar to functional programming languages such as Gofer or ML. For example:

```
data SensorVal = Defect | Busy | Ready(Float);
```

defines the type `SensorVal` (within a package), together with the constants `Defect`, `Busy` and a constructor function `Ready : Float -> SensorVal`. In addition, a partial selector function `ReadySel1 : SensorVal -> Float` is implicitly defined by the data declaration. Furthermore, three discriminator functions: `is_Defect, is_Busy, is_Ready : SensorVal -> Bool` are defined implicitly as well.

Additional user-specific functions and constants can be defined within the DTDs, as for instance

```
fun nextValue(last,Defect) = last
  | nextValue(last,Busy) = last
  | nextValue(last,Ready(x)) = x;
```

All definitions, constructor and selector functions, discriminator predicates, and user-specific functions of such a DTD are generated into an Ada package. Each generated Ada package, as usual, is realized by an Ada specification file declaring the publicly available functions of the package and an implementation (body) file implementing these functions based on private (non-publicly accessible) implementations of the data types. This principle of maximum encapsulation for the code generation applies not only to the aforementioned data types, but also to all other AutoFocus model elements and views.

AutoFocus SSDs describe the system structure and the communication flow between the components (channels). Each SSD could be mapped into an Ada package with types for each subsystem. However, for efficiency reasons we decided to flatten the system into atomic components and generate a simple Ada package encompassing the system as a whole and one Ada package for each type of component. Communication (i.e., data flow between component ports along channels) is implemented by a simple copy function between the interface ports of the atomic components. A port can also be empty (if no value was written to it in the current, respectively, previous clock cycle), so with each port a "data present" flag is associated.

AutoFocus STDs describe the behavior of components. They are similarly encoded into packages with a type for each component including a state variable, declarations of ports and local variables. Furthermore there is a main procedure ("`Do_Transition`") that executes the component (one transition on each invocation). This method represents the semantics: according to the current state, all possible transitions are checked and the first executable transition is executed. Since this function can be quite complex (depending on the number of states and transitions in the STD), for each state and for each transition separate helper functions can be generated (switched on or off in the generator settings) to reduce the McCabe complexity of the generated code, however, for the price of an increase in the number of generated modules.

## 5.3 Transformation of Model Tests to Program Tests

Now that we have briefly sketched how models are transformed into Ada code and thus into an implementation in a programming language, we still have to describe how test cases on the model level (test sequences, given by EETs) are transformed to the program level. For that purpose, we first have a look at such an EET (Fig. 8), generated by a simulation of the LES model.

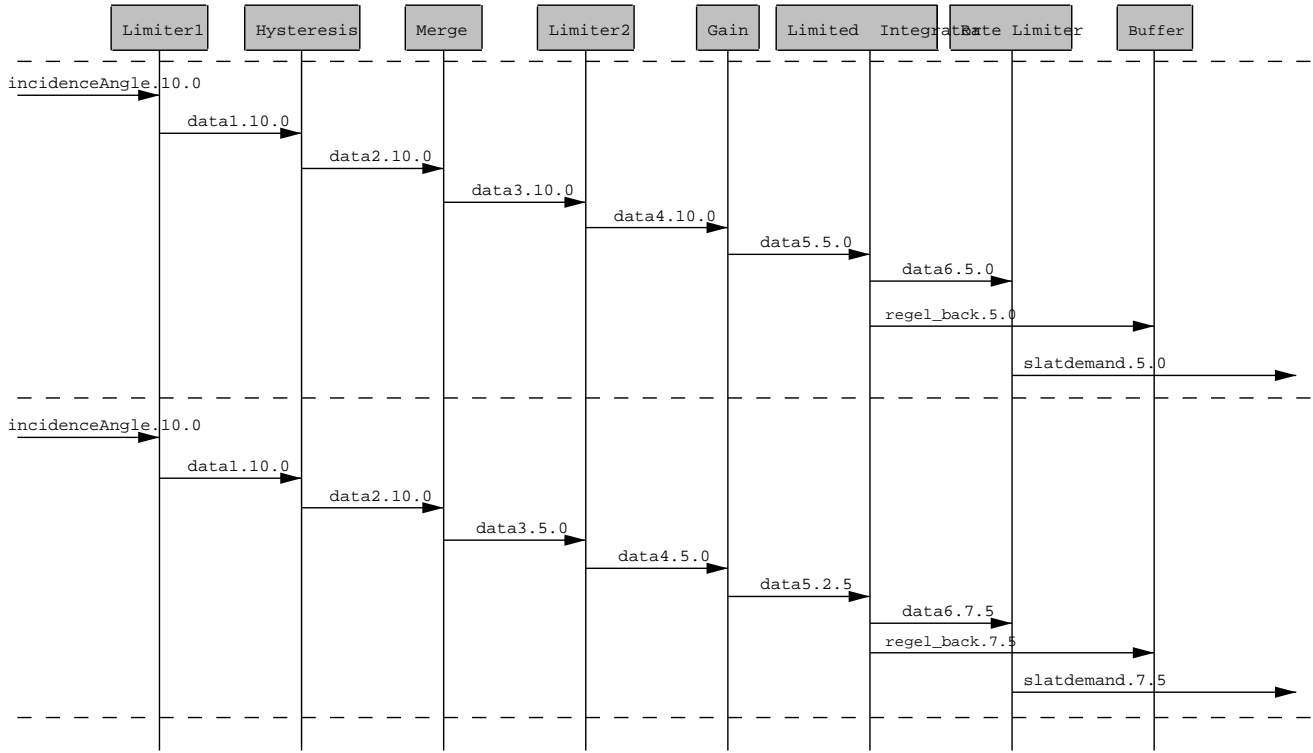

Fig. 8: Part of a Model Test Case/Sequence (EET) from a LES Model Simulation

A test sequence given by such an EET is actually a test case on the model level for each of the components involved in the EET. When creating a test sequence for a *single one* of these components, we first have to isolate the component and the communication events it receives and sends from the rest of the EET. This is basically done by a projection on the respective component's axis in the EET. This single component test sequence can now be transformed into a test for the generated code, where each of the clock cycles (delimited by dashed horizontal lines) in the EET, including the incoming and outgoing communication events, corresponds to an invocation of the main `Do_Transition` procedure in the generated Ada code. Thus, a model-based test sequence delivers not only one, but a complete series of test cases for the program code.

### 5.3.1 Relationships between Code Generation and Test Case Transformation

The relationship between the code generation mechanism and the transformation of model tests into program tests is actually quite close. Not only does each test sequence correspond to a series of program tests, even further, this relationship involves two important assumptions about the internal state of the components and their Ada-based implementation as well:

First, an EET records a communication (input/output) *sequence* for the component involved, which is well-sorted with respect to the system clock cycles. Since an EET does not reveal the internal state (control state and data state) of the component ("black box view"), and since each input/output event in the sequence is a result not only of the received input, but also of the current internal component state, it is mandatory that the sequence of clock cycles is preserved when executing the resulting set of program test cases on the generated code.

Second, an EET obviously has a starting point, the first clock cycle, where all communication begins. It is our common interpretation (although differing interpretations are as well admissible) that the internal component state at this starting point of an EET is the initial state (both for control state and for internal data values) of the component.

Only with these two rules, a well-defined transformation of an EET test sequence into a series of program tests is possible.

It is also worth mentioning that we are dealing with fundamentally different notions with respect to unit-, subsystem-, and system-testing on the model level and on the program level: In the model world, the units of interest for testing (or for test case generation) are, quite naturally, components. Therefore, an isolated axis from an EET with input-/output events for only one component can be regarded as a unit test for that component. As previously (Section 5.2) described, however, a model unit (i.e., component) is implemented by numerous program constructs in the generated Ada code, ranging from the packages that encapsulate the component and state space properties down to atomic functions and procedures that implement, for instance, a predicate for testing whether a given transition is enabled or not, or a procedure that actually carries out all the state changes effected by that transition. Thus, what is conceptually seen as a unit test on the model level, actually expands to a whole subsystem- or integration-test on the program code level. Nevertheless, it is as well possible, using the data from the model-based test cases, to drive unit tests on the code level: If, for instance, only one specific transition predicate function is of interest for testing, it is possible—although not implemented at this time and currently not planned to be realized—to scale down the generated program tests for that purpose.

### 5.3.2 Test Execution and Analysis

To actually let developers execute the described testing scheme using our tools, two different approaches are available at this time, and a third one is currently being implemented. Fig. 9 gives a schematic overview of these approaches.

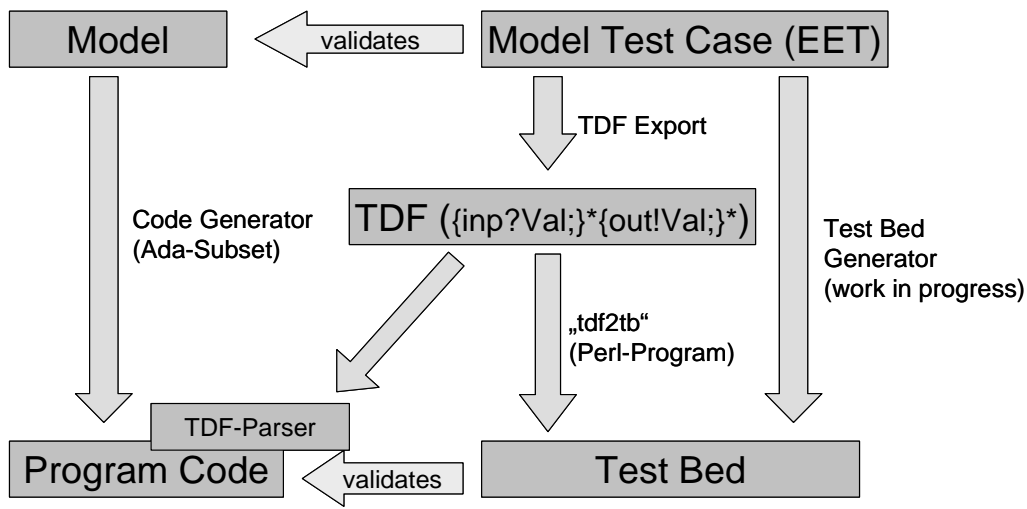

Fig. 9: Available Options for Test Execution on Generated Ada Code

Similarly to the model, which is transformed into an Ada-based implementation by the code generator, model test cases are transferred into program tests. The first two mentioned approaches are based on exporting an EET into an intermediate text format, called TDF (short for Test Data Format). TDF files contain one line for each execution step (i.e., clock cycle) of the system; this line consists of input and output patterns (similar to patterns of transitions in STDs) that associate data values with the component ports that they are read on (in case of input ports) or written to (in case of output ports):

```
{inp?val;}*{outp!val;}*
```

TDF-based testing of a system can be done in two different ways: First, by directly feeding a TDF file into the TDF parser, which can be generated as an add-on to the Ada code by the code generator. This way, the generated Ada has its own, integrated testing engine. Second, a TDF file can be directly converted into an Ada program that acts as a test bed for the component of interest in the generated Ada code. This is done by "tdf2tb", a small Perl program implemented in MOBASIS. The third way to transform a model test into a test bed program is currently being implemented; it directly converts the relevant parts of an EET into an Ada test bed, without the necessity of using the TDF intermediate format.

In any of the three cases, the actual testing of the program code is done by setting the input values of a component's ports, letting the component execute a transition, and comparing the output ports with the required values.

Now that we do execute tests on actual program code (and no longer on models), we can actually use standard Ada testing tools to assess the quality of our test cases with respect to structural criteria as the attained coverage level of the code, for instance in terms of MC/DC, which is mandatory for risk class 1 software systems. Within the MOBASIS project, we have chosen the Rational Test RealTime toolset (formerly: ATTOL UniTest & Coverage) for that purpose. Thus, by instrumenting all or parts of the generated code to measure the desired coverage criteria, we can immediately—after completion of a test run—obtain a report indicating the coverage level that has been achieved by our—originally model-based—test case.

## 6 Combined Testing Process

So far, we have shown the way from models to Ada code and from model tests to program tests. Although the direction of activity always points from the model level to the program level this approach is by no means to be a one-way process.

### 6.1 Iterated Testing Cycles: "Roundtrip"

The results of executing the transformed model-based tests on the generated program code have two aspects. First, a functional one: If the model-based tests delivered functionally correct results on the model level, it is obvious that their transformation to the program level should deliver identical results on the generated program code (if not, then this would indicate errors either in the code generator or in the test case transformation tool). Testing both on the model level and on the program level may seem unnecessary and redundant at first sight, however, the benefits of such a combination will become clearer soon. The second aspect of the test execution on the generated program code is a structural one: As mentioned before, an instrumentation of the code can deliver valuable data concerning the coverage on the program code that a model-based test case achieves. With these results that indicate, for instance, which parts of the code have not sufficiently been covered by the test, we can—due to the directly traceable mapping between model elements and generated code—easily analyze our model for the insufficiently covered parts and subsequently modify, e.g., the boundaries for a constraint-based search algorithm to deliver structurally better test cases. With the new test cases obtained in this way, we can again perform testing on the code level and observe the hopefully improved coverage results.

Repeating this process several times leads to a *roundtrip test case optimization process* that uses automated techniques to a large extent both on the model and on the code level, and therefore promises to deliver high-quality test cases (both from a functional and structural point of view) rapidly.

## 6.2 Benefits of the Combined Approach

An approach, as just described, that combines model-based development and test case generation with code-based text execution and analysis has several benefits for developers using this approach and the associated tools, as well as for the method and tool developers themselves.

Model-based development uses domain-oriented abstractions that do not contain implementation-related information to an extent as found in programming languages. Models are therefore independent from any particular programming language and can be re-used as a whole or in parts for projects that will in the end be implemented in different programming languages (the availability of suitable code generators is obviously a prerequisite for this).

Since models are more compact and abstract than programs, they are more amenable to validation techniques than programs, such as model checking or constraint-based searches. Model tests (in the form of EETs) can be created (and archived for later usage on the code) by prototyping/simulation and other validation techniques very early in the development process even before the first line of code of a system has been created. For these reasons, it is more promising to perform the often tedious task of finding test cases on models rather than on programs.

Since, however, the accepted measures for test coverage that are required in the development of safety-critical software systems are defined only for programs and not for models at this time (defining such measures for a specific modeling language is a research topic on its own) it is only consequent to transfer both model an model tests to the code level, where the required measures are available. This is particularly obvious when the actual system implementation is to be carried out by code generation from models. Thus, in such a setting, it is possible to combine the particular strengths of the model-based world with those of the programming language world.

What has not been explicitly mentioned so far, is the issue of correctness of code generators and test case transformation tools (which is another area of research within MOBASIS). Proving the correctness of such tools by formal means is—although possible in theory—practically impossible due to the effort required. Therefore, in practice often the principle of trust in tools and tool developers by continuous usage of tools without any major quality problems replaces formal proofs of correctness. In this context, the seemingly redundant possibility to execute tests both on models and on generated code can help the tool developers in finding remaining errors in their code generators (in case of inconsistent test results), and thus in continuously augmenting the quality of their tools and—with a growing record of projects with consistent results—in establishing confidence in the quality of their tools.

## 7 Conclusion and Further Work

In this paper, we have given a comprehensive overview of some of the activities in the MOBASIS project related to the fields of testing, test case generation, and code generation. We have introduced the AutoFocus modeling framework, have highlighted some of the applied principles in model-based test case generation, and have then shown the basics of the generation of Ada code from models and of the transformation of model tests into program tests, all of these illustrated by parts of the MOBASIS example application, the leading edge system.

We have pointed out a combined approach that promises to help developers in tackling one of the most challenging activities during development, in finding a set of test cases for their implementation that provides relatively optimal coverage results. As described, the sketched approach combines the specific strengths of model-based and code-based development and can thus be regarded as a "Best-of-both-Worlds" approach. Its evaluation that is currently taking place will show how and to what extent it will in fact help developers in carrying out this task.

Although the project-related implementation activities will be finished by the end of the project, many activities, as indicated, are still considered work in progress, and, in particular, the fundamental research activities in the field of test case generation will remain an area of active research even far beyond the end of MOBASIS.

# 8 References

1. J. Barnes: High Integrity Ada, The SPARK Approach. Praxis Critical Systems Ltd., Addison-Wesley, 1997.
2. G. Booch, I. Jacobson, J. Rumbaugh: The Unified Modeling Language User Guide. Addison-Wesley, 1998.
3. M. Broy, O. Slotosch: Enriching the Software Development Process by Formal Methods. Proceedings of FM-Trends 98, LNCS 1641.
4. F. Huber, S. Molterer, A. Rausch, B. Schätz, M. Sihling, O. Slotosch: Tool Supported Specification and Simulation of Distributed Systems. Proceedings of International Symposium on Software Engineering for Parallel and Distributed Systems, 1998.
5. F. Huber, B. Schätz: Integrated Development of Embedded Systems with AutoFOCUS. Technical Report TUM-I0107, Fakultät für Informatik, TU München, 2001.
6. International Telecommunication Union: Message Sequence Charts, 1996. ITU-T Recommendation Z.129, Geneva, 1996.
7. J. Philipps, O. Slotosch: The Quest for Correct Systems: Model Checking of Diagrams and Datatypes. Proceedings of Asia Pacific Software Engineering Conference 1999, pp. 449-458.
8. Pretschner: Classical Search Strategies for Test Case Generation with Constraint Logic Programming. Proc. Formal Approaches to Testing, pp. 47-60, 2001.
9. RTCA Inc., EUROCAE: Software Considerations in Airborne Systems and Equipment Certification. DO-178B / ED-12B, 1992.
10. O. Slotosch: Quest: Overview over the Project. Proceedings of FM-Trends 98, 1998 LNCS 1641:346-350.
11. N. Tracey: A Search-Based Automated Test-Data Generation Framework for Safety-Critical Software. PhD thesis, University of York, 2000.
12. G. Wimmel, A. Pretschner, O. Slotosch: Specification Based Test Sequence Generation with Propositional Logic. Journal on Software Testing Verification and Reliability (to appear).